\documentclass[final.1p]{elsarticle}

\usepackage{amssymb}
\usepackage{amsmath}
\usepackage{graphicx}
\usepackage{subfigure}
\usepackage{subcaption}
\usepackage{indentfirst}
\usepackage[switch]{lineno}
\usepackage{color}
\usepackage[colorlinks=true, allcolors=blue]{hyperref}
\usepackage{float}
\usepackage{stfloats}
\usepackage{longtable}
\usepackage{titlesec}

\usepackage{tikz}
\usepackage{circuitikz}

\hypersetup{colorlinks = true,linkcolor = blue,anchorcolor =red,citecolor = blue,filecolor = red,urlcolor = red,
            pdfauthor=author}

\journal{}

\begin{document}

\begin{frontmatter}



\title{High precision proton beam monitor system concept design on CSNS based on SiC}
\author[NJU,IHEP]{Ye He}
\author[Jilin]{Xingchen Li}
\author[IHEP]{Zijun Xu}
\author[NJU]{Ming Qi}
\author[IHEP]{Congcong Wang}
\author[IHEP]{Chengwei Wang}
\author[NJU]{Hai Lu}
\author[CSNS]{Xiaojun Nie}
\author[CSNS]{Ruirui Fan}
\author[CSNS]{Hantao Jing}
\author[Jilin]{Weiming Song}
\author[IHEP,Liaoning]{Keqi Wang}
\author[LZU1,LZU2]{Kai Liu}
\author[IHEP]{Peilian Liu}
\author[IHEP,Tsinghua]{Hui Li}
\author[IHEP]{Zaiyi Li}
\author[IHEP]{Chenxi Fu}
\author[IHEP]{Xiyuan Zhang}
\author[Liaoning]{Xiaoshen Kang}
\author[IHEP]{Zhan Li}
\author[IAT]{Suyu Xiao\corref{cor2}}
\ead{suyu.xiao@iat.cn}
\author[IHEP]{Xin Shi\corref{cor1}}
\ead{shixin@ihep.ac.cn}

\address[NJU]{Nanjing University, Gulou District, Nanjing 210093, People’s Republic of China}
\address[IHEP]{Institute of High Energy Physics, Chinese Academy of Sciences, Shijingshan District, Beijing 100049, People’s Republic of China}
\address[Jilin]{Jilin University, Chaoyang District, Changchun 130015, People’s Republic of China}
\address[CSNS]{Spallation Neutron Source Science Center, People’s Republic of China}
\address[Liaoning]{Liaoning University, Huanggu District, Shenyang 110136, People’s Republic of China}
\address[LZU1]{School of Nuclear Science and Technology, Lanzhou University, Lanzhou 730000, People’s Republic of China}
\address[LZU2]{Frontiers Science Center for Rare Isotopes, Lanzhou University, Lanzhou 730000, People’s Republic of China}
\address[Tsinghua]{Tsinghua University, Haidian District, Beijing 100084, People’s Republic of China}
\address[IAT]{Shandong Institute of Advanced Technology, Licheng District, Jinan 250000, People’s Republic of China}
\cortext[cor1,cor2]{Corresponding authors}

\begin{abstract}
A high precision beam monitor system based on silicon carbide PIN sensor is designed for China Spallation Neutron Source 1.6 GeV proton beam to monitor the proton beam fluence.
The concept design of the beam monitor system is finished together with front-end electronics with silicon carbide PIN sensors, readout system and mechanical system.
Several tests are performed to study the performance of each component of the system.
The charge collection of the SiC PIN sensors after proton radiation is studied with 80 MeV proton beam for continuous running. Research on the performance of the front-end electronics and readout system is finished for better data acquisition.
The uncertainty of proton beam fluence is below 1$\%$ in the beam monitor system. 

\end{abstract}




\end{frontmatter}

\section{Introduction}
China Spallation Neutron Source (CSNS) \cite{ref-CSNS,ref-CSNS2} consists of white neutron source \cite{ref-neutron}, Associated Proton beam Experiment Platform (APEP), proton test beam facility and muon beam facility. The white neutron source and APEP have been built, while others are still in progress. A beam monitor system is necessary to detect the fluence of proton beam. Diamond is commonly used is ATLAS \cite{ref-diamond} and CMS \cite{ref-design1} for beam monitor, while silicon carbide sensors are used in beam monitors for its good performance in radiation hardness and timing resolution \cite{ref-review,ref-sic-application} compared to diamond sensors. Studies have shown the uncertainty of beam fluence measured by beam monitor based on semiconductor detector is about 10 $\%$ \cite{ref-accuracy}. Several methods are taken to improve the precision and reduce the uncertainty to 1$\%$ at 1 Hz counting rate in this work. 

The preliminary design of the beam monitor system is finished which consists of detector board, pre-amplifier, and readout system. The detector board is based on 5 mm $\times$ 5 mm SiC PIN sensors \cite{ref-PIN-beta} fabricated by Nanjing University (NJU). Radiation impact on leakage current, capacitance and charge collection efficiency of the SiC PIN sensors are studied and the results prove that the performance of the SiC PIN meets the requirements of the beam monitor system. 8 movable SiC PINs are used around the beam cross section on a mechanical system. A pre-amplifier board is optimised according to UCSC single channel board \cite{ref-board} by changing the shape of the board to satisfy spatial constraints around the beam line. The readout system is designed for the signals from front-end electronics and calculate the beam fluence.

\section{Design of the beam monitor system}
The beam monitor system consists of pre-amplifier, readout system and mechanical design \cite{ref-design2}. The system is based on 5 mm $\times$ 5 mm SiC PIN sensors fabricated by NJU to detect the proton beam. Fig. \ref{fig1}(a) shows the position of detectors around the beam line. Each SiC PIN sensor is equipped with motor to tune the position of the SiC PIN sensors to proper location. 

\subsection{Beam parameter}
\begin{figure}[h]
    \centering
    \begin{minipage}[t]{0.48\textwidth}
        \centering
        \includegraphics[width=0.8\textwidth]{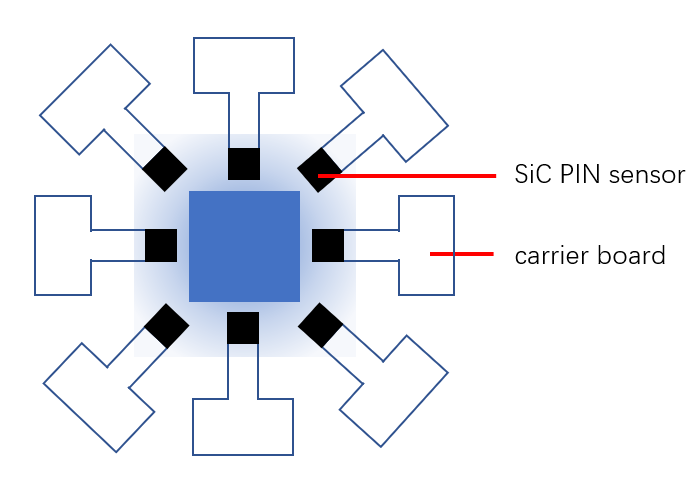}
        \captionsetup{labelfont={bf},name={},labelsep=period}
        \caption*{(a)}
    \end{minipage}
    \begin{minipage}[t]{0.48\textwidth}
        \centering
        \includegraphics[width=0.8\textwidth]{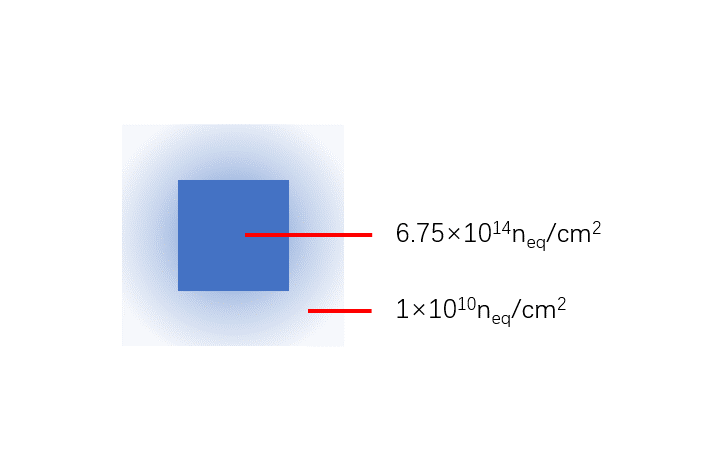}
        \captionsetup{labelfont={bf},name={},labelsep=period}
        \caption*{(b)}
    \end{minipage}
    \caption{(a)The layout of the detector. It consists of 8 movable SiC PIN sensors around the beam center. (b)The profile and fluence of the beam.\label{fig1}}
\end{figure}

The energy of the proton beam is designed to be 1.6 GeV \cite{ref-CSNS2} in CSNS upgrade, and fluence is $2.35$×$10^{8}~\rm n_{eq}/cm^{2}$. Radiation dose is 6.75 × $10^{14}~\rm n_{eq}/cm^{2}$ at center and 1 × $10^{10}~\rm n_{eq}/(cm^{2}\times s^{-1})$ at edge as shown in Fig. \ref{fig1}(b). The beam fluence is uniformly distributed in the beam center and decreases at edge. The beam fluence can be deduced by the signal from SiC PIN sensors and it is estimated from the beam fluence that about each 20 ns comes a proton hitting the SiC PIN sensor at edge of beam.

\subsection{Carrier board}
The pre-amplifier is a transimpedance amplifier \cite{ref-board}, mounted on a T-shape carrier board. Charge collection efficiency of the board is tested by 355 nm laser and $^{90}$Sr source. The results are shown in section 4.

\subsection{Readout system}
The readout system is composed of CAEN DT 5742 \cite{ref-DT5742}, data line and control system. DT 5742 can collect waveform in 200 ns. It starts sampling and generate a waveform when a proton hits the SiC PIN sensor, and then the data will be sent to the control system. A software built inside the system is used to analyze the data and get the number of protons.

\section{Performance of the SiC PIN}
\subsection{Electrical performance}
\begin{figure}[h!]
    \centering
    \begin{minipage}[t]{0.48\textwidth}
        \centering
        \includegraphics[width=0.96\textwidth]{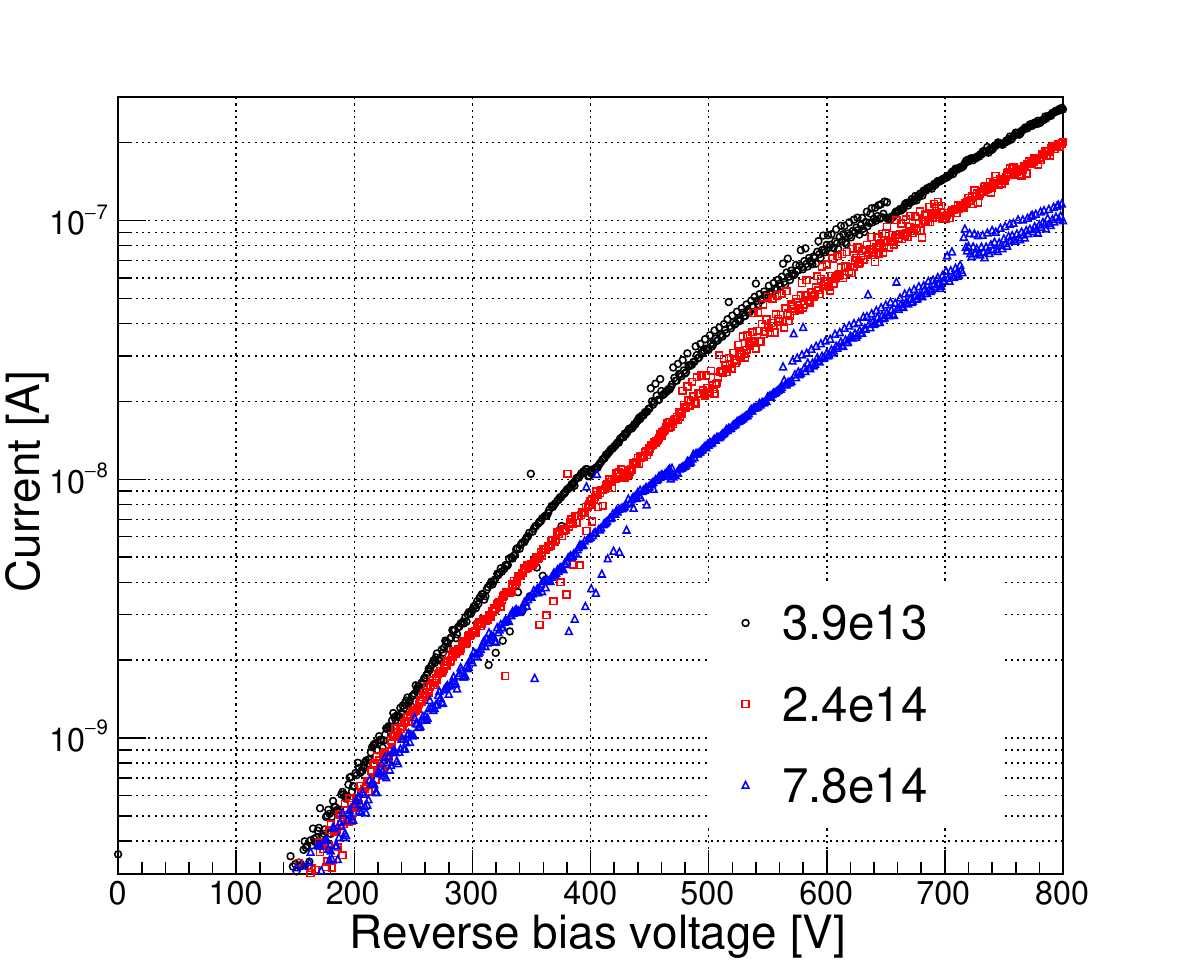}
        \captionsetup{labelfont={bf},name={},labelsep=period}
        \caption*{(a)}
    \end{minipage}
    \begin{minipage}[t]{0.48\textwidth}
        \centering
        \includegraphics[width=0.96\textwidth]{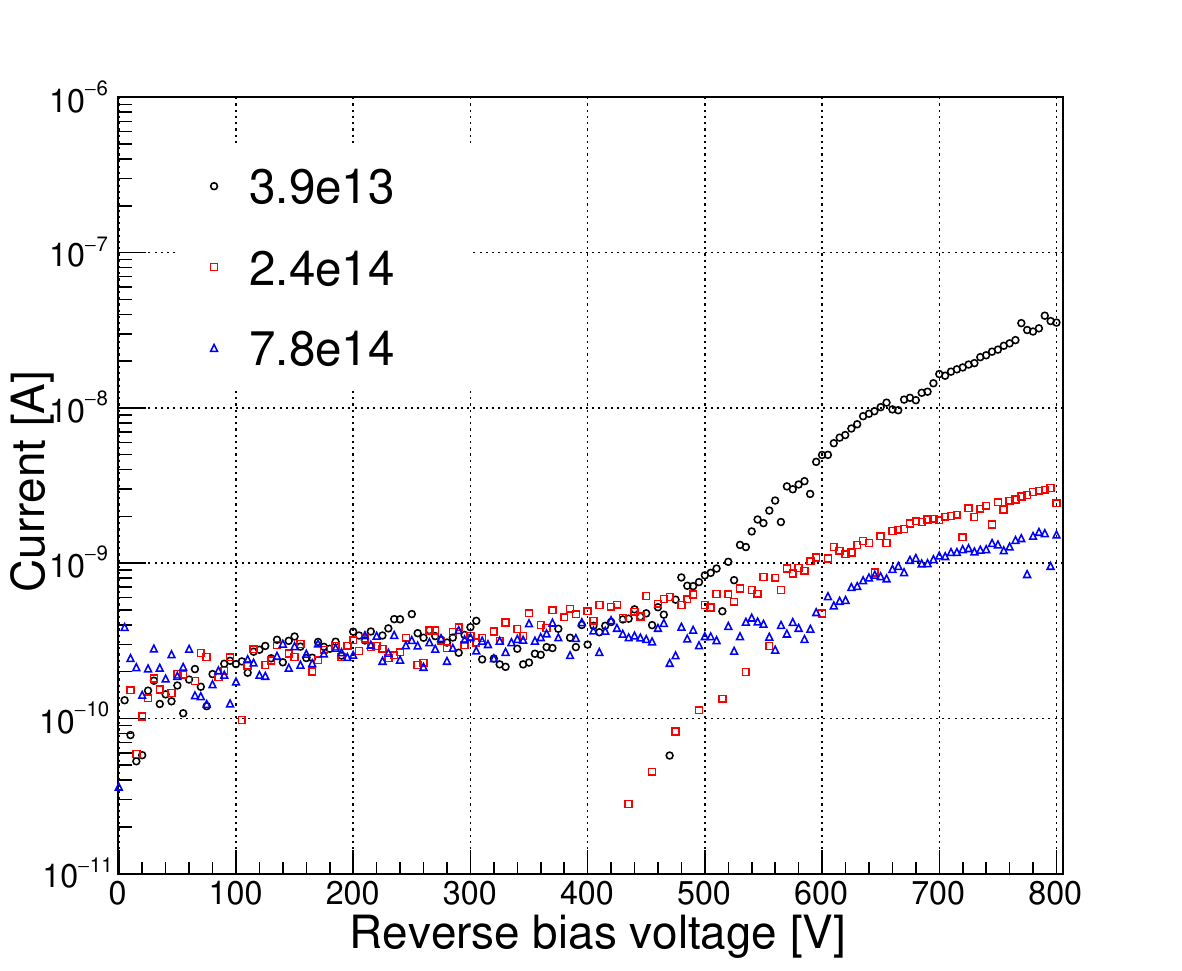}
        \captionsetup{labelfont={bf},name={},labelsep=period}
        \caption*{(b)}
    \end{minipage}
    \caption{Current-voltage characteristics (a) before radiation and (b) after radiation\label{fig2}}
\end{figure}

\begin{figure}[h!]
    \centering
    \begin{minipage}[t]{0.48\textwidth}
        \centering
        \includegraphics[width=0.96\textwidth]{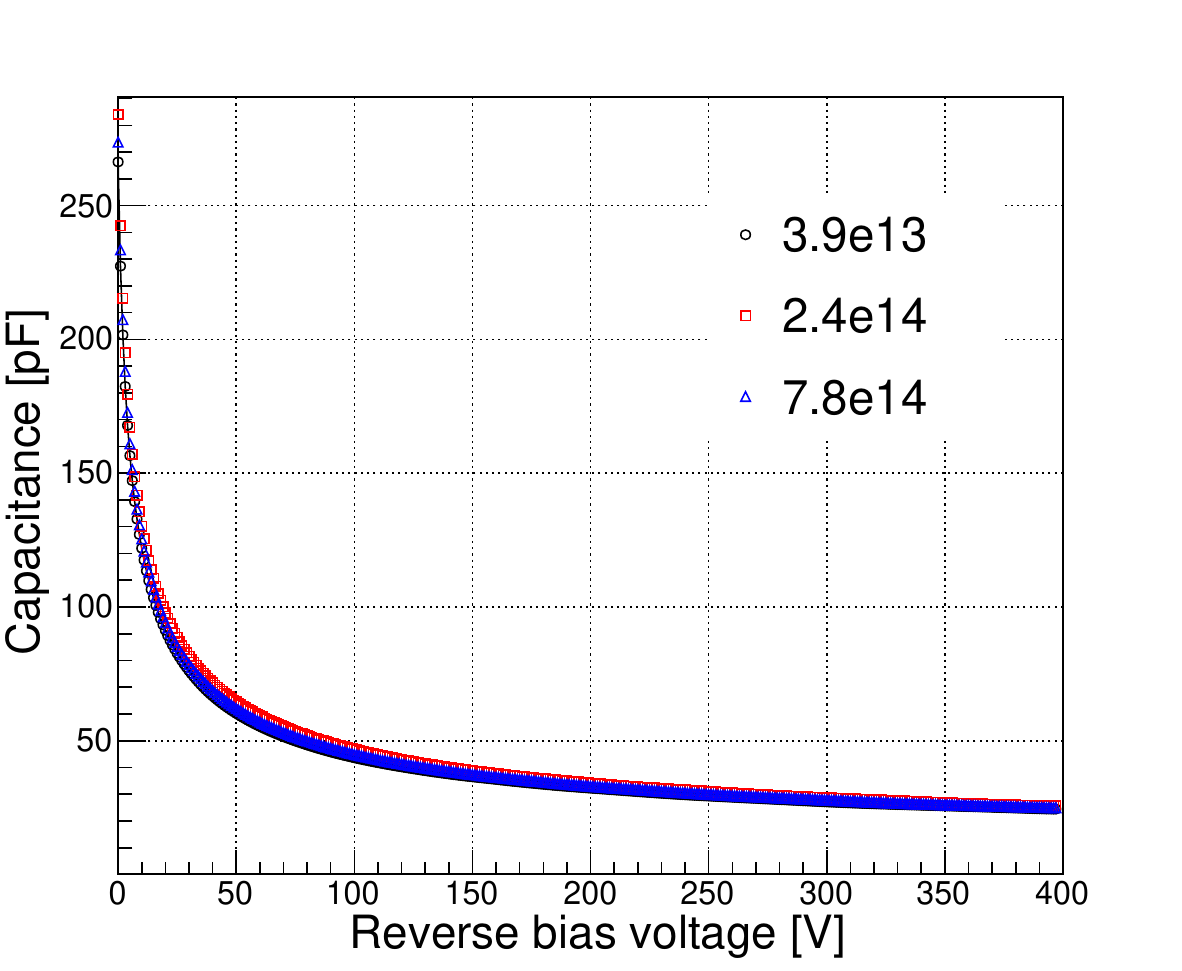}
        \captionsetup{labelfont={bf},name={},labelsep=period}
        \caption*{(a)}
    \end{minipage}
    \begin{minipage}[t]{0.48\textwidth}
        \centering
        \includegraphics[width=0.96\textwidth]{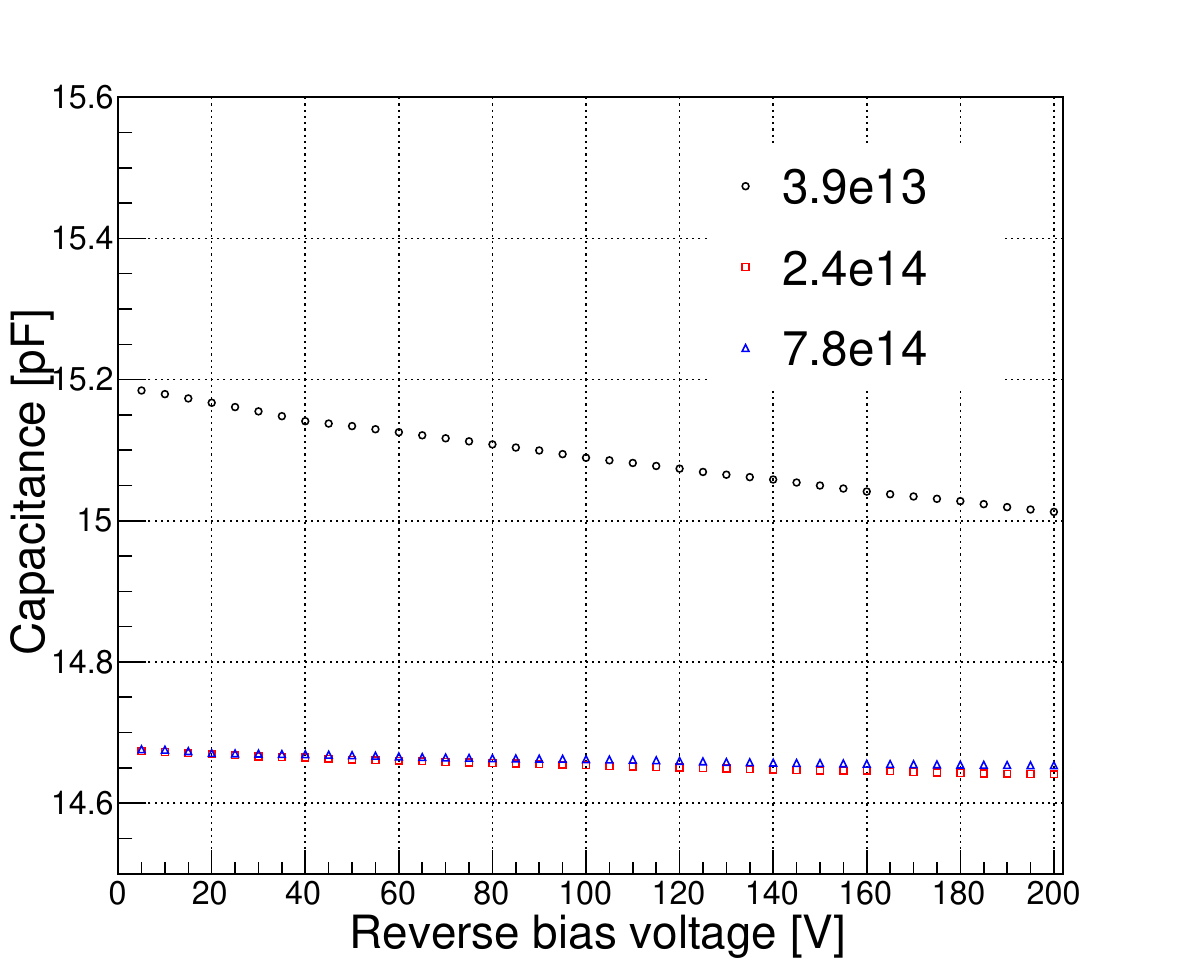}
        \captionsetup{labelfont={bf},name={},labelsep=period}
        \caption*{(b)}
    \end{minipage}
    \caption{Capacitance-voltage characteristics (a) before radiation and (b) after radiation\label{fig3}}
\end{figure}

The performance of the SiC PIN sensor after proton radiation \cite{ref-irradiation1} is studied by carrying out radiation test on them. The radiation test is performed at APEP and three points are tested, PIN1$-3.9\times10^{13}~\rm n_{eq}/cm^{2}$, PIN2$-2.4\times10^{14}~\rm  n_{eq}/cm^{2}$, PIN3$-7.8\times10^{13}~\rm n_{eq}/cm^{2}$, where $\rm n_{eq}/cm^{2}$ is the 1 MeV neutron equivalent radiation dose per square centimeter. Fig. \ref{fig2} shows the current-voltage characteristics of the SiC PIN sensor before and after radiation. IV characteristics of three SiC PIN sensors are similar before radiation and leakage current becomes smaller after radiation. It is indicated in Fig. \ref{fig3} that the CV characteristics of the SiC PIN sensor before radiation are similar to each other. Capacitance decreases after radiation and it becomes smaller as radiation dose increases \cite{ref-irradiation1,ref-irradiation2}. It is revealed in Fig. \ref{fig2} and Fig. \ref{fig3} that performance of the SiC PIN sensors will degrade after radiation.

\begin{figure}[h!]
    \centering
    \includegraphics[width=0.48\textwidth]{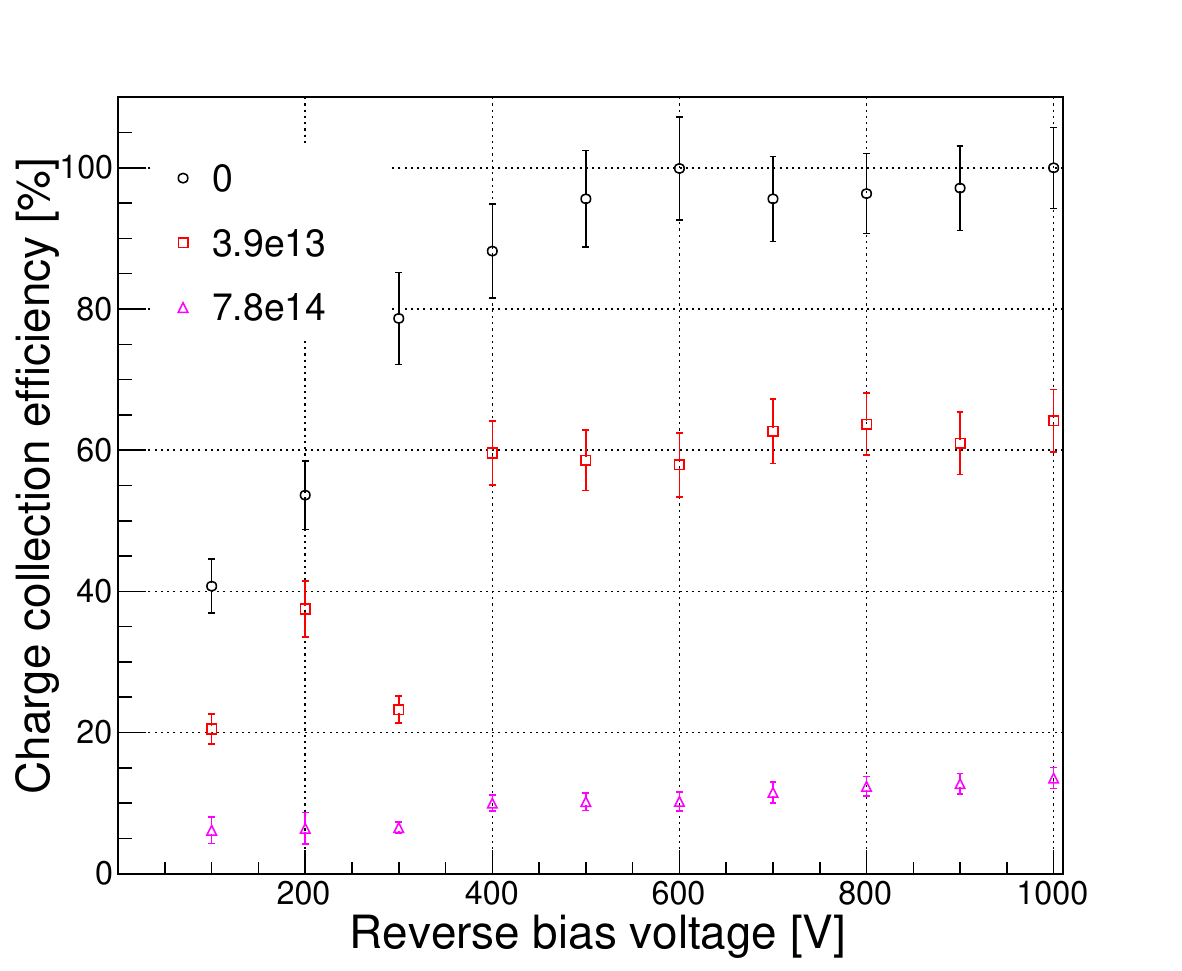}
    \caption{Charge collection efficiency of the SiC PIN sensor before and after radiation.\label{fig4}}
    \label{fig:my_label}
\end{figure}

The charge collection efficiency (CCE) is measured by 355 nm laser \cite{ref-CCE} whose power is 1.44$\times 10^{-6} \rm~W$ and the result is shown in Fig. \ref{fig4}. The efficiency of unirradiated sensor at 600V is selected to be 100$\%$. It is indicated from Fig. \ref{fig4} that charge collection efficiency remains the same when the SiC PIN sensor is full depleted as it can be seen in CV characteristic Fig. \ref{fig3}. Detecting efficiency drops to 60$\%$ at 500 V after $3.9\times10^{13}~\rm n_{eq}/cm^{2}$ proton radiation. The efficiency becomes smaller as radiation dose gets larger and at $7.8\times10^{14}~\rm n_{eq}/cm^{2}$ the efficiency is only about 10$\%$. 

\subsection{Calibration}
The charge collection efficiency after radiation can be calibrated according to radiation dose. Fig. \ref{fig3} indicates that the efficiency drops to about 60$\%$ at 500V at 7.8$\times 10^{14}~\rm n_{eq}/cm^{2}$. Assuming that the change in charge collection varies linearly \cite{ref-CCE2} at small radiation dose around $1\times 10^{13}~\rm n_{eq}/cm^{2}$, the relationship between time and efficiency can be calculated. $7.8\times 10^{14}~\rm n_{eq}/cm^{2}$  is not concluded in calibration because extreme high dose like this locates out of the linear range. The relationship between the charge efficiency $\eta$ and Q is described by the following relationship
\begin{equation}
    \eta = (1 - \frac{Q}{3.9\times 10^{13}}\times 0.6)\times 100\% , Q<3.9\times 10^{13}
\end{equation}

Radiation dose of each SiC PIN sensor is about 1 × $10^{12}$ and charge collection efficiency is 98\% according to the relationship.

\section{Performance of readout board}
\subsection{USCS single channel board}
The performance of UCSC single channel board \cite{ref-board} is studied to apply the board to the beam monitor system. It is a transimpedence amplifier (TIA), and its gain is 20 dB. The feedback resistance is the main electric component affecting amplification factor. Simulations and $\rm \beta$ source tests are performed to study its performance.

\subsection{NGspice simulation}
\begin{figure}[!h]
    \centering
    \begin{minipage}[t]{0.48\textwidth}
        \centering
        \includegraphics[width=0.96\textwidth]{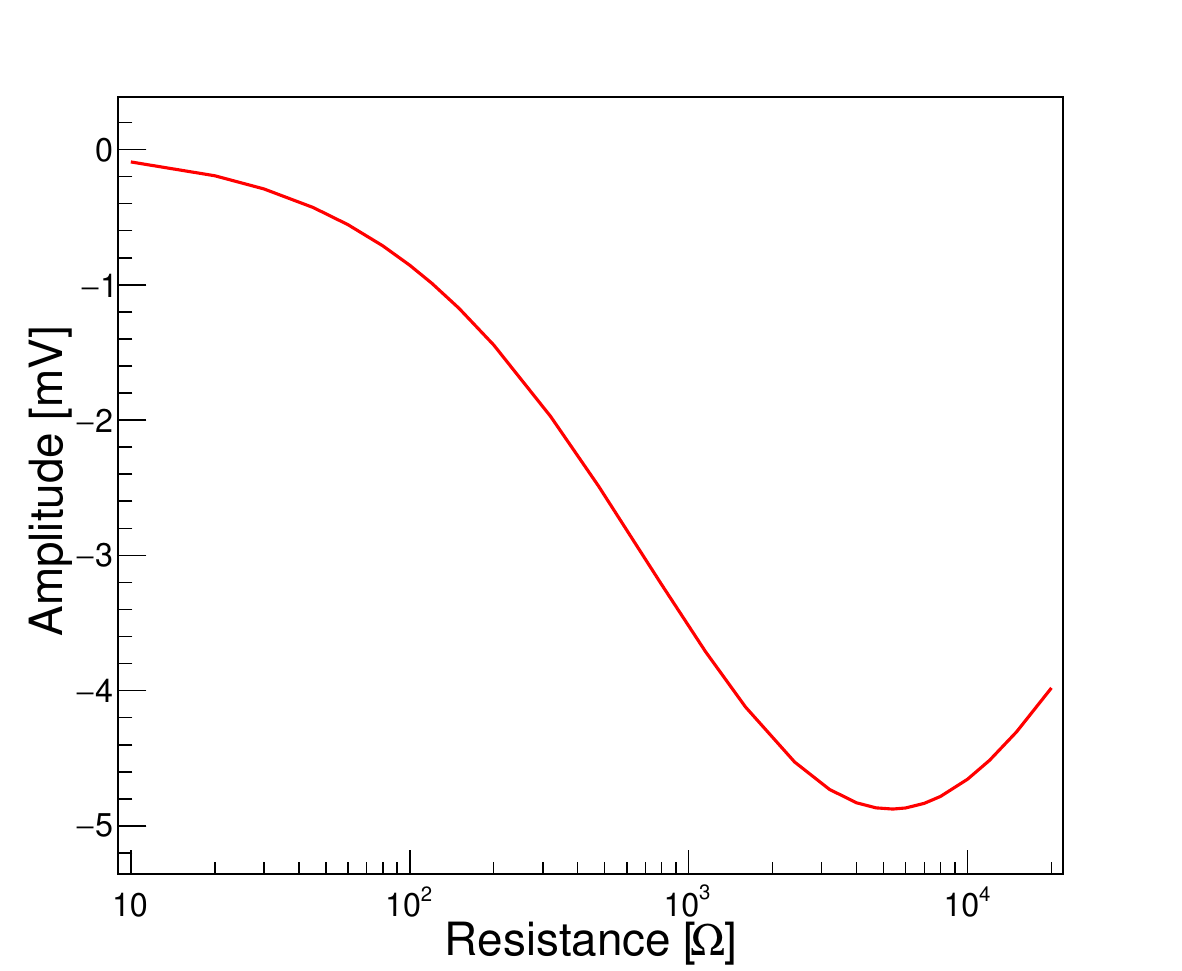}
        \captionsetup{labelfont={bf},name={},labelsep=period}
        \caption*{(a)}
    \end{minipage}
    \begin{minipage}[t]{0.48\textwidth}
        \centering
        \includegraphics[width=0.96\textwidth]{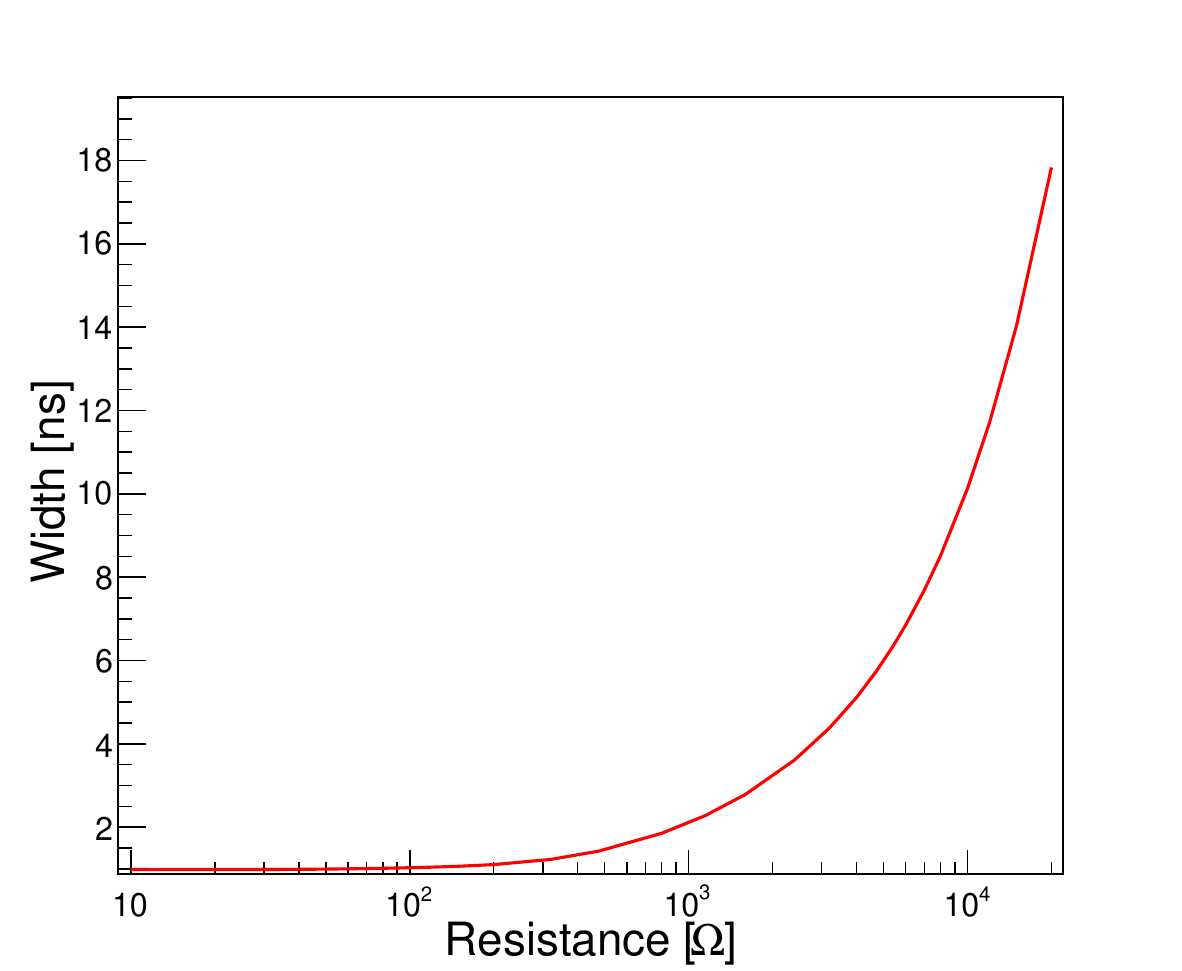}
        \captionsetup{labelfont={bf},name={},labelsep=period}
        \caption*{(b)}
    \end{minipage}
    \caption{NGspice simulation result of the performance of pre-amplifier when the resistance of the feedback resistor changes. (a) shows the relationship between resistance of the feedback resistor and the amplitude of output signal. (b) shows the relationship between resistance of the feedback resistor and the width of output signal. \label{fig5}}
\end{figure}

NGspice \cite{ref-NGspice} is used to simulate the performance of the pre-amplifier circuit. The results of simulation are shown in Fig. \ref{fig5} by adjusting the value of the feedback resistor. Gain of the board reaches an maximum with a 4000 $\Omega$ feedback resistor as Fig. \ref{fig5}(a) shows, which is what we expect. The result indicates that board with 4000 $\Omega$ feedback resistor has the best performance in amplifying.

\subsection{\texorpdfstring{$\beta$}{} source test on SiC PIN}
\begin{figure}[!h]
    \centering
    \begin{minipage}[t]{0.48\textwidth}
        \centering
        \includegraphics[width=0.96\textwidth]{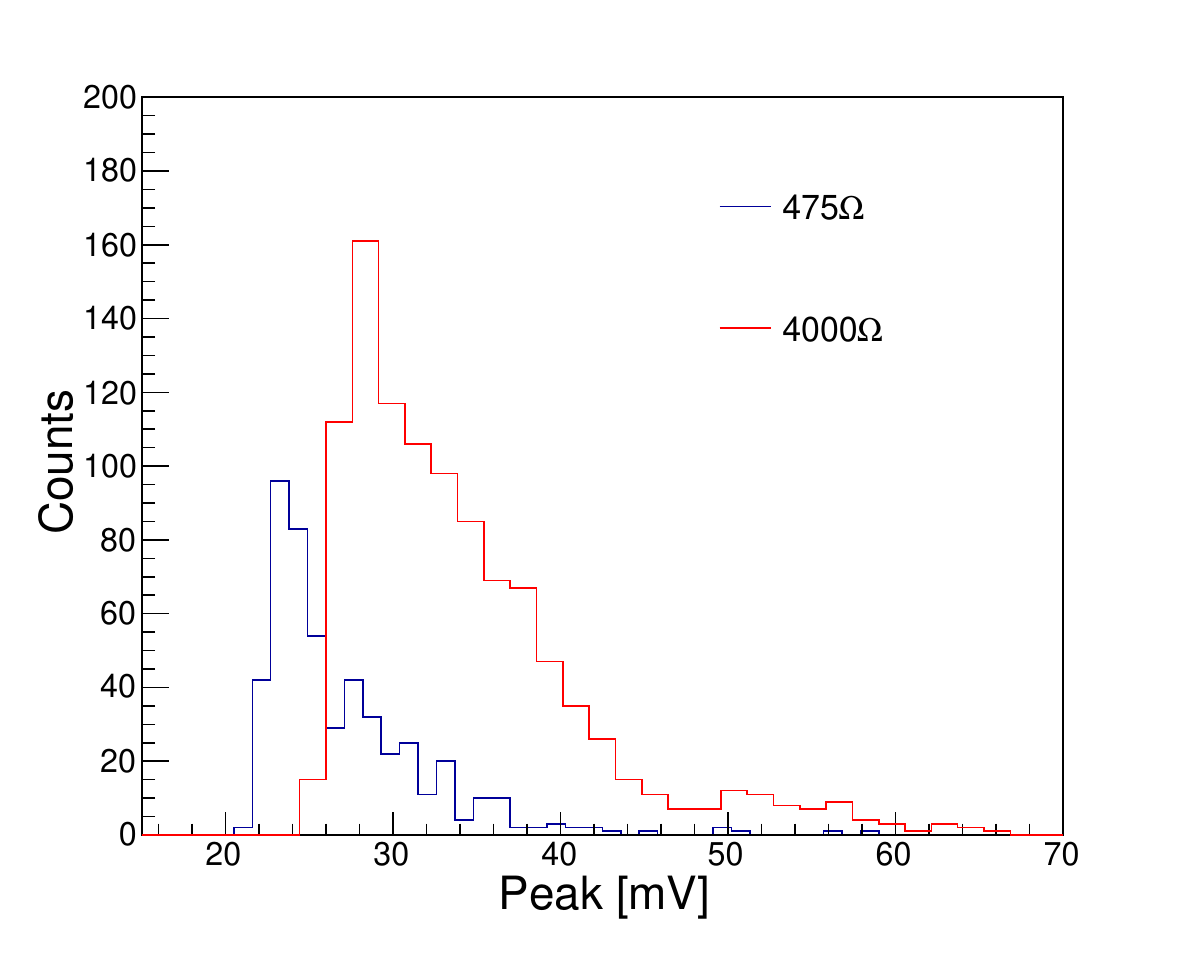}
        \captionsetup{labelfont={bf},name={},labelsep=period}
        \caption*{(a)}
    \end{minipage}
    \begin{minipage}[t]{0.48\textwidth}
        \centering
        \includegraphics[width=0.96\textwidth]{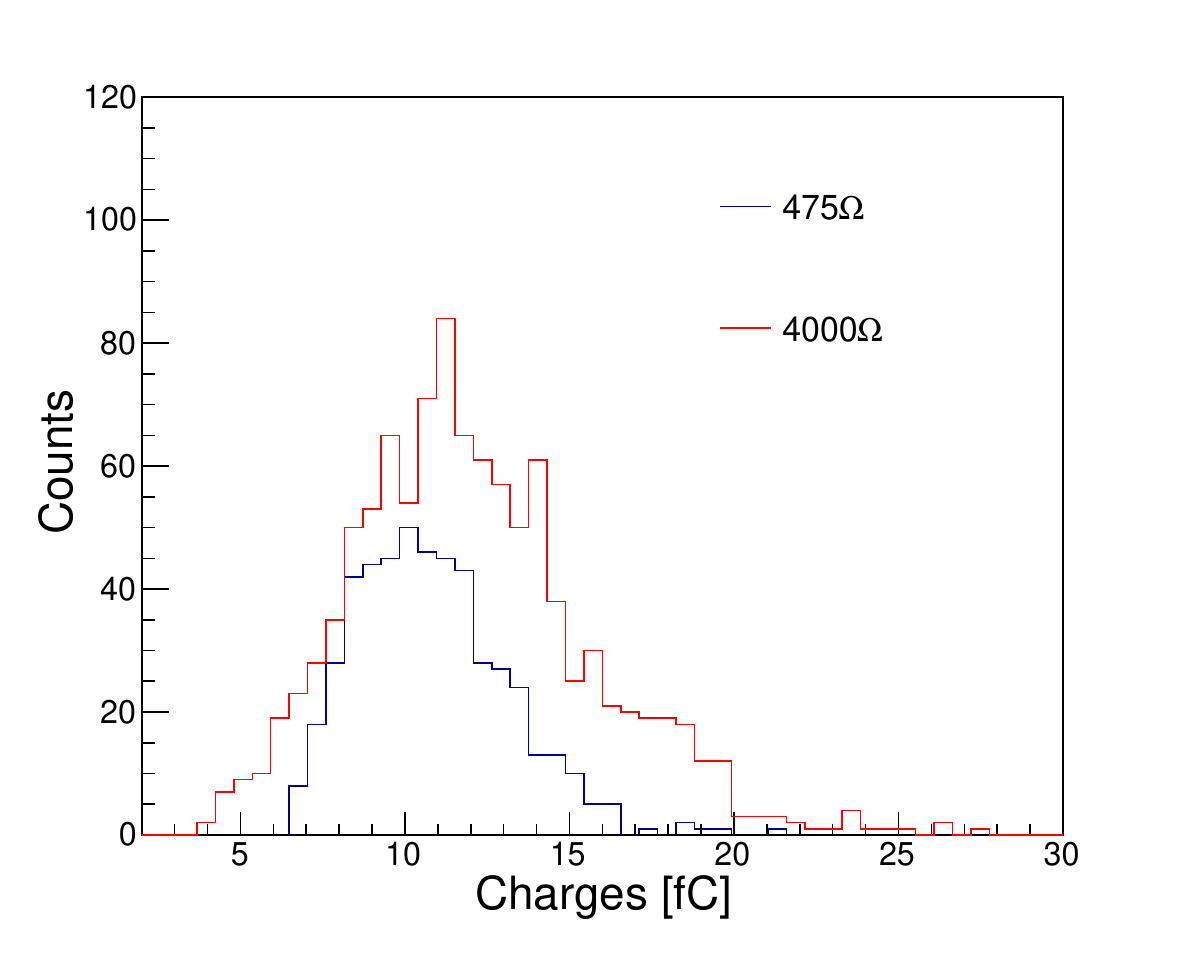}
        \captionsetup{labelfont={bf},name={},labelsep=period}
        \caption*{(b)}
    \end{minipage}
    \caption{The result of $\beta$ source test on the UCSC single channel board with 475 $\Omega$ feedback resistor and 4000 $\Omega$ feedback resistor. (a) shows the distribution of signal peaks, (b) shows the distribution of charge. \label{fig6}}
\end{figure}

$^{90}$Sr $\beta$ source tests are finished with UCSC board before and after changing the feedback resistor. The source meter is Keithley 2470 and SiC PIN sensor is supplied with 500 V reverse bias voltage. The output signal is connected to a 20 dB main amplifier PE1513. The trigger voltage is set to be 21 mV. Results are shown in Fig. \ref{fig6}. It is indicated in Fig. \ref{fig6} that board with $4000~\Omega$ feedback resistor is able to detect electrons in a larger charge range than $475~\Omega$ feedback resistor and the average peak voltage is larger, which means more electrons can be detected. Therefore 4000 $\Omega$ is selected to be the feedback resistor. 

\section{Systematic uncertainties}

The uncertainty of detecting beam fluence can be divided into Landau noise $\sigma_{\rm Landau}$, electronic noise $\sigma_{\rm elec}$, noise from proton radiation effects $\sigma_{\rm rad}$ and beam fluctuation $\sigma_{\rm fluc}$. Uncertainty in this system is defined as signal loss $\rm (N_{loss}/N_{signal})\times 100\%$. The total uncertainty is $\sigma_{\rm total}$.

\begin{equation}
    \sigma_{\rm total} = \sqrt{\sigma_{\rm Landau}^{2} + \sigma_{\rm elec}^{2} + \sigma_{\rm rad}^{2} +
    \sigma_{\rm fluc}^{2}}
\end{equation}

\subsection{Landau noise}
\begin{figure}[h]
    \centering
    \includegraphics[width=0.48\textwidth]{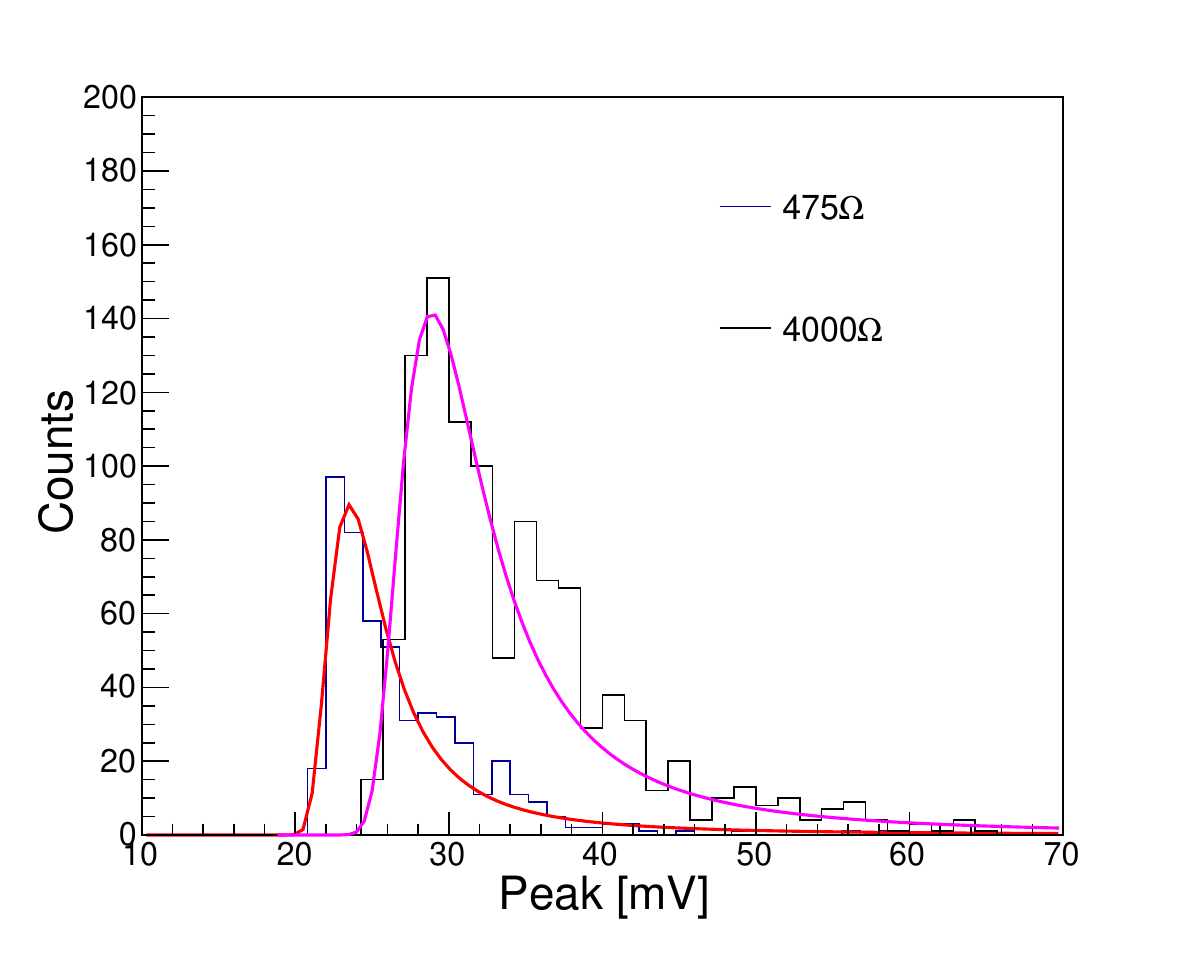}
    \caption{Landau fitting in peak voltage distribution of $\beta$ source test when the feedback resistor is 475$ \Omega$ and 4000 $\Omega$. \label{fig7}}
\end{figure}

The $\sigma_{\rm Landau}$ comes from the Landau distribution in energy deposition of electrons \cite{ref-Landau}. Fig. \ref{fig7} indicates results of Landau fitting.

\begin{figure}[h!]
    \centering
    \includegraphics[width=0.48\textwidth]{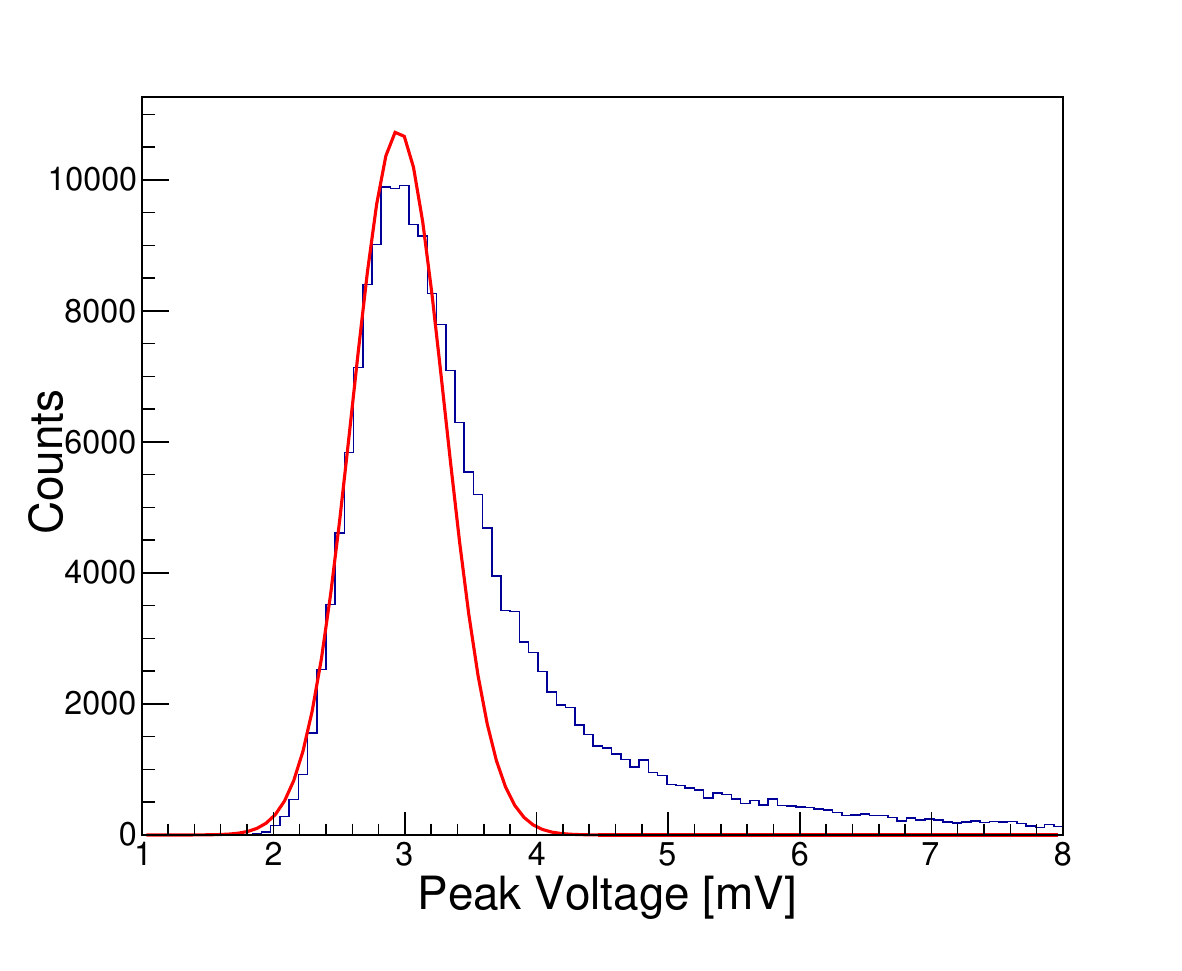}
    \caption{Peak voltage distribution of signal generated from SiC detectors in simulation and Gaussian fitting on peak voltage. The fitting is good at peak voltage smaller than 3 mV. $\sigma_{
    \rm G}$is 0.351 mV and mean peak voltage $\mu_{\rm peak~voltage}$ is 2.950 mV.}
    \label{fig8}
\end{figure}

The results reveals that more electrons can be detected when the feedback resistor is 4000 $\Omega$ and it results in smaller uncertainty. RASER \cite{ref-raser} is used to simulate the proton beam and signals from the SiC detector are shown in Fig. \ref{fig8}. Gaussian fitting is used to estimate uncertainty.

Fig. \ref{fig8} shows Gaussian fitting on peak voltage and fitting curve is good at peak voltage smaller than 3 mV, so it is reasonable to use to calculate the distribution of peak voltage less than 2 mV. $\sigma_{
\rm G}$is 0.351 mV and mean peak voltage $\mu_{\rm V}$ is 2.950 mV. $\sigma_{\rm L}$ is 0.263 mV if Landau fitting is used. Using Gaussian fitting instead of Landau fitting brings larger uncertainty, for $\sigma_{\rm G}$ is larger than $\sigma_{\rm L}$. Therefore, the uncertainty calculated from Gaussian distribution is larger than Landau distribution, which means uncertainty $\sigma_{\rm Landau}$ is smaller than the distribution of peak voltage less than 2 mV. Let V be the peak voltage, the distribution of peak voltage less than 2 mV $\rm N_{V\leq2~mV}$ can be calculated by the following equation:
\begin{equation}
    \rm N_{V\leq2~mV}=\int_{0}^{2}{\frac{1}{\sqrt{2\pi}\sigma_{G}}e^{-\frac{(V-\mu_{peak})^2}{2\sigma_{G}^2}}}dV
\end{equation}
The result $\rm N_{V\leq2~mV}$ is 0.33\%, and this means $\sigma_{\rm Landau}$ is 0.33\% at most. The maximum value of uncertainty $\sigma_{\rm Landau}$ is taken, and $\sigma_{\rm Landau}$ is 0.33\%.

\subsection{Electronic noise}

The uncertainty of electronic noise $\sigma_{\rm elec}$ comes from the board and leakage current of SiC PIN sensors. It is indicated in Fig. \ref{fig9} that noise is mainly distributed at below 1.2 mV. The distribution of noise amplitudes indicates that mean value is 0.57 mV $\sigma_{\rm noise} \rm ~is~ 0.18$ mV, so it is important to set a suitable trigger to reduce noise and reserve more true signals. If the trigger is 2 mV, less than 0.3$\%$ noise signal will be detected, so $\sigma_{\rm elec}\rm ~is~0.3\%$. 

\begin{figure}[h!]
    \centering
    \includegraphics[width=0.48\textwidth]{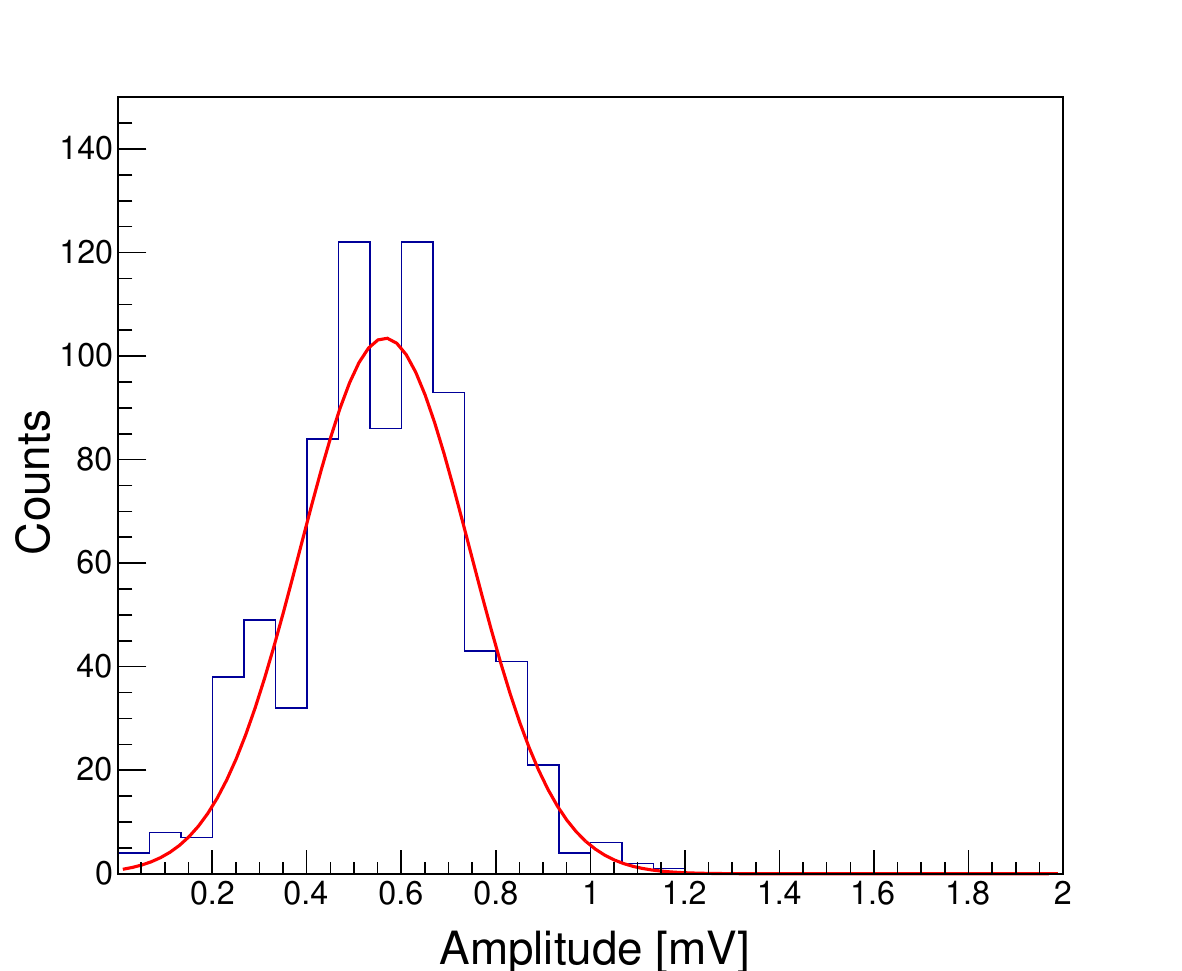}
    \caption{Output noise distribution of UCSC single channel board\label{fig9}}
    \label{fig9}
\end{figure}

\subsection{Radiation}
The uncertainty of radiation $\sigma_{\rm rad}$ comes from the decrease in performance when the SiC PIN sensor is under proton radiation. Charge collection efficiency will drop to about 98$\%$ as calculated in section 3.2 if the system runs 5000 hours in a year. Less than $0.1\%$ signals will be lost at 2 mV trigger, and the signal is less than $0.1\%$ which means $\sigma_{\rm rad}\rm ~is~0.1\%$.

\subsection{Fluctuation}
The uncertainty of beam fluctuation $\sigma_{\rm fluc}$ comes from shifting of the beam profile. It is necessary to figure out the relationship between beam fluctuation and fluence to correct the fluctuation for the fluctuation will affect detecting precision. It is assumed that beam fluence is uniformly distributed at center($r<1~\rm cm$) and is a Gaussian distribution from $r=1~\rm cm$ to $r=3~\rm cm$, and then is evenly distributed at $3~\rm cm<r<4~\rm cm$ and $r>4~\rm cm$, where the fluence has a sharp drop at $r=4~\rm cm$, the distribution is shown in the following equation :
\begin{equation}
    f(r) = \left\{
    \begin{aligned}
    & 1 \times 10^{9}\ ,\ 0 < r < 1~\rm cm \\
    & 4.52\times 10^{9}\times\frac{1}{\sqrt{2\pi}\times 0.761}\times exp(-\frac{r^2}{2\times 0.761^{2}})\ ,\
    1~\rm cm < r < 3~\rm cm \\
    & 1 \times 10^{6}\ ,\ 3~\rm cm < r < 4~\rm cm \\
    & 3 \times 10^{4}\ ,\ r > 4~\rm cm \\
    \end{aligned}
    \right.
\end{equation}

\begin{figure}[h]
    \centering
    \includegraphics[width=0.48\textwidth]{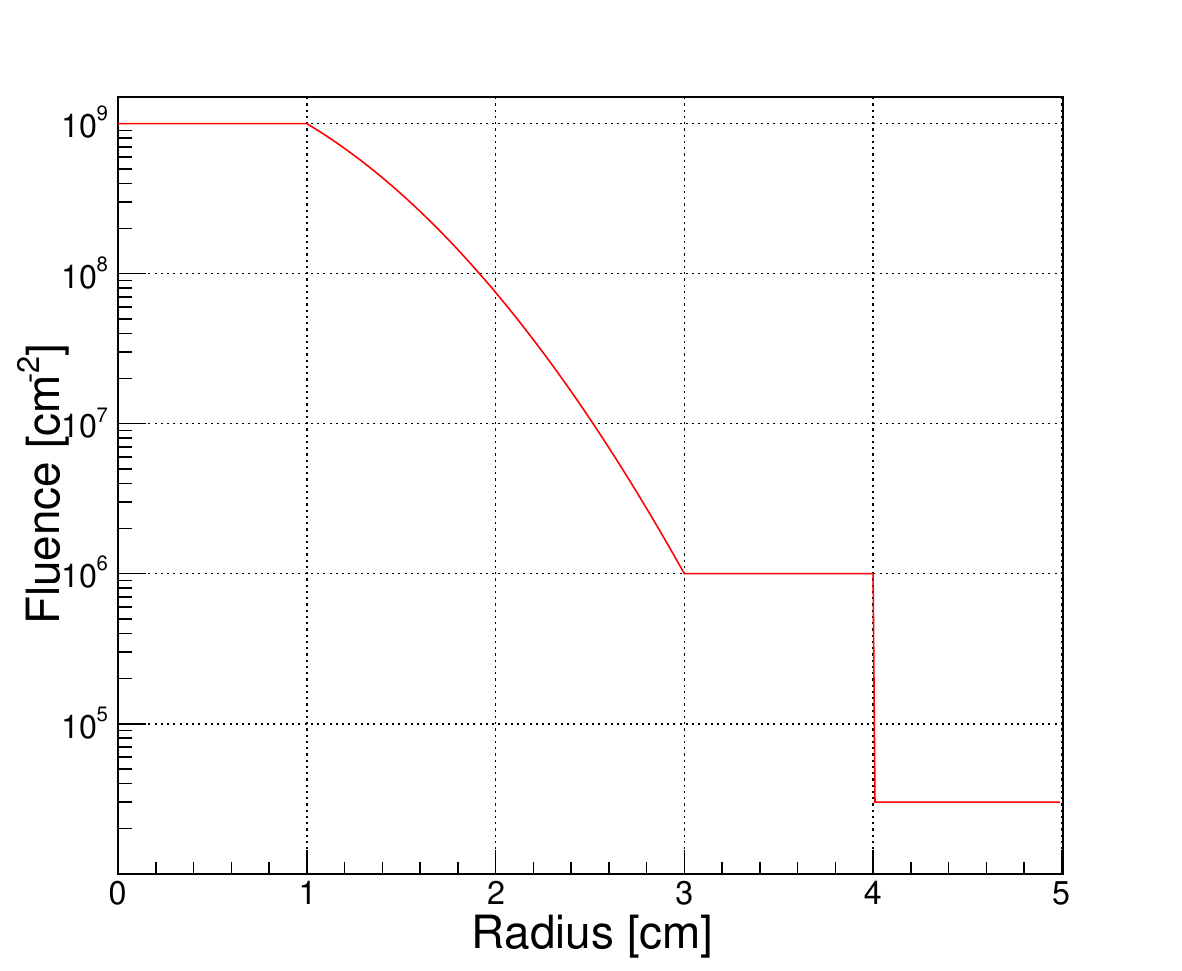}
    \caption{The relationship between radius and fluence. It is assumed that beam fluence is uniformly distributed at center which is a square of 2 cm $\times$ 2 cm and gradually decreases till the edge. Therefore the fluence is even at $r<1~\rm cm$ and is a Gaussian distribution from $r=1~\rm cm$ to $r=3~\rm cm$, and then is uniformly distributed at $3~\rm cm<r<4~\rm cm$ and $r>4~\rm cm$.}
    \label{fig10}
\end{figure}

\begin{figure}[h!]
    \centering
    \includegraphics[width=0.48\textwidth]{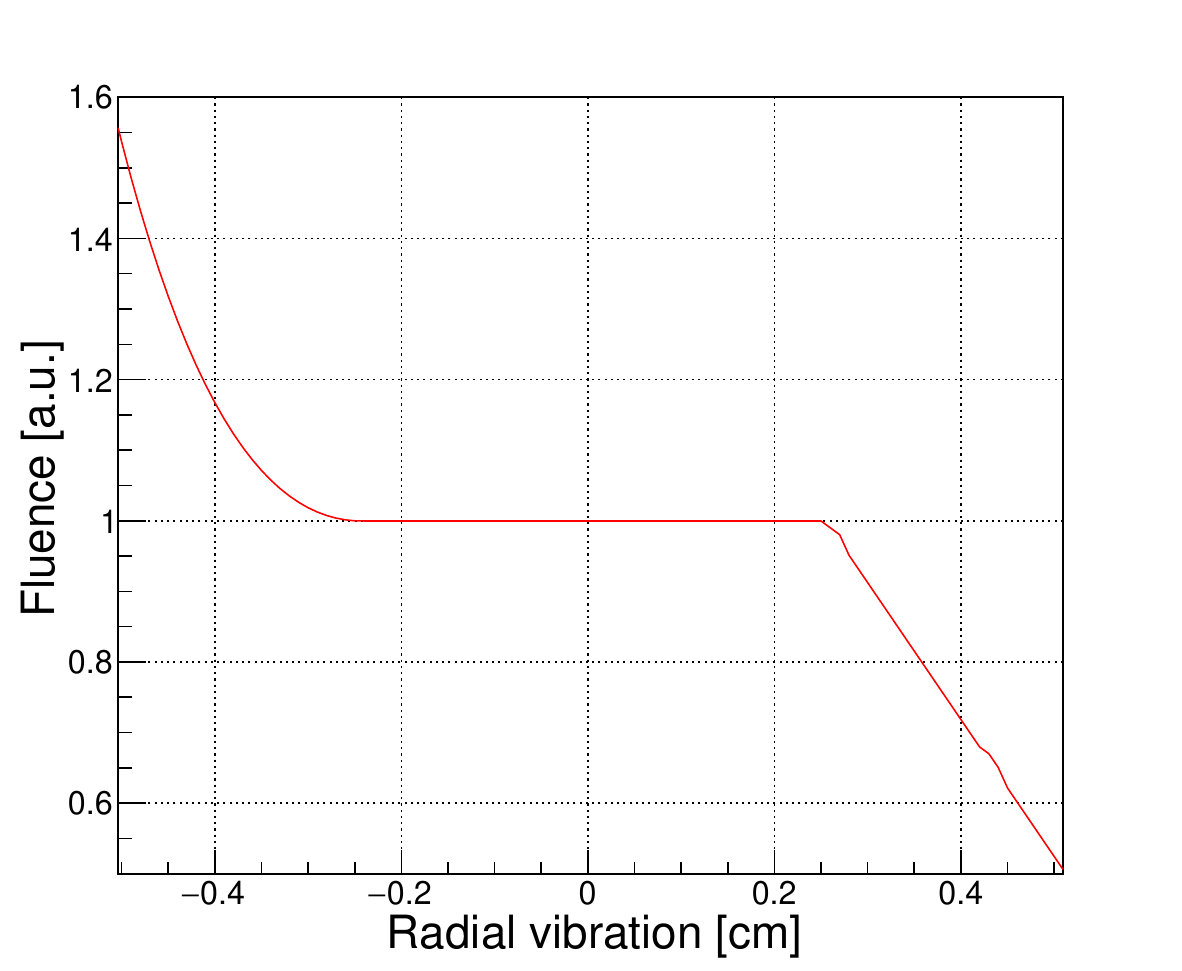}
    \caption{Relationship between radial fluctuation and fluence}
    \label{fig11}
\end{figure}

The distribution of the beam is shown in Fig. \ref{fig10}. Detector board is set at $r=3.5~\rm cm$, where the fluence is smaller than center and the total fluence will be $1 \times 10^{12}/\rm cm^{2}$. Fig. \ref{fig11} shows the relationship between relative fluence of single SiC PIN sensor against initial fluence. The fluence will not change if the radial beam fluctuation is in range $\pm 0.25~\rm cm$, and radial fluctuation can be calculated out when the fluctuation is in range of $\pm 0.25~\rm cm$ to move the SiC PIN sensor to suitable location. Because the beam fluence varies linearly in $1~\rm cm<r<3~\rm cm$ and has a rapid drop at $r=4~\rm cm$, the fluence also varies linearly when the radial fluctuation is greater than $0.25~\rm cm$. 

The SiC PIN sensor carrier board is designed to be movable and can be controlled by motor system, so $\sigma_{\rm fluc}$ comes from the uncertainty of the motor system $\Delta_{\rm motor}$. Assuming that $\Delta_{\rm motor}=50$ $\mu$m, the $\Delta_{\rm fluence}$ will be $0.05\%$ from Fig. \ref{fig11}. So the $\sigma_{\rm fluc}\rm ~is~0.50\%$ and it requires $\Delta_{\rm motor} \leq \rm 50$ $\mu$m.

\subsection{Uncertainty}
Considering all the factor in Table \ref{tab1}, the final uncertainty will be
\begin{equation}
    \sigma_{\rm total} = \sqrt{\sigma_{\rm Landau}^{2} + \sigma_{\rm elec}^{2} + \sigma_{\rm rad}^{2} + \sigma_{\rm fluc}^{2}} = 0.68\%
\end{equation}

\begin{table}[H]
    \centering
    \caption{Uncertainty in beam monitor system}
    \begin{tabular}{ll}
    \hline
         Source  &$\sigma$ / \% \\
         \hline
         $\sigma_{\rm Landau}$  &0.33 \\
         $\sigma_{\rm elec}$ &0.30  \\
         $\sigma_{\rm rad}$ &0.10 \\
         $\sigma_{\rm fluc}$ &0.50 \\
         \hline
    \end{tabular}
    \label{tab1}
\end{table}
The uncertainty is $0.68\%$, which can meet the goal of $1\%$, so the design of the beam monitor system is able to satisfy requirements in CSNS proton beam monitor project.

\section{Conclusion}
A beam monitor system is designed for CSNS 1.6 GeV proton beam in its next upgrade. Preliminary design of beam monitor system is raised in this work. The system is composed of SiC PIN sensor, carrier board, mechanical system and readout system. Performance of the sensor after radiation is studied and the results prove its good performance in radiation hardness. The single channel board is redesigned based on UCSC readout board to carry the SiC PIN sensor and electrical tests on the board performed to prove the board is suitable for the beam monitor system. The uncertainty of the system is estimated to be below $1\%$.

The beam monitor system can be used to perform more studies for radiation with the low uncertainty. Performance of sensors after radiation can be studied with precise radiation dose. Furthermore, it is possible to expand the system to other beamlines just by modifying the carrier board and readout system in the future.

\section*{Acknowledgement}
This work is supported by the National Natural Science Foundation of China (No. 12305207, No. 11961141014, No. 12375184, No. 12205321), the State Key Laboratory of Particle Detection and Electronics (No. SKLPDE-ZZ-202312, No. SKLPDE-KF-202313), the Natural Science Foundation of Shandong Province Youth Fund (No. ZR202111120161).

\end{document}